\documentclass{article}

\usepackage{microtype}
\usepackage{graphicx}
\usepackage{subfigure}
\usepackage{booktabs} 

\usepackage{hyperref}

\usepackage[accepted]{icml2023}

\usepackage{amsmath}
\usepackage{amssymb}
\usepackage{mathtools}
\usepackage{amsthm}

\usepackage[capitalize,noabbrev]{cleveref}

\theoremstyle{plain}

\theoremstyle{definition}

\theoremstyle{remark}

\usepackage[textsize=tiny]{todonotes}

\icmltitlerunning{Weak Chemical Tagging with Graph Kernel Method}

\begin{document}

\twocolumn[
\icmltitle{Weisfeiler-Lehman Graph Kernel Method: \\ A New Approach to Weak Chemical Tagging}



\icmlsetsymbol{equal}{*}

\begin{icmlauthorlist}
\icmlauthor{Yuan-Sen Ting}{equal,rsaa,soco,osu}
\icmlauthor{Bhavesh Sharma}{equal,soco}
\end{icmlauthorlist}

\icmlaffiliation{soco}{School of Computing, Australian National University, Acton, ACT 2601, Australia}
\icmlaffiliation{rsaa}{Research School of Astronomy \& Astrophysics, Australian National University, Cotter Rd., Weston, ACT 2611, Australia}
\icmlaffiliation{osu}{Department of Astronomy, The Ohio State University, Columbus, USA}

\icmlcorrespondingauthor{Yuan-Sen Ting}{yuan-sen.ting@anu.edu.au}
\icmlcorrespondingauthor{Bhavesh Sharma}{Bhavesh.Sharma@anu.edu.au}

\icmlkeywords{Machine Learning, ICML}

\vskip 0.3in
]


\printAffiliationsAndNotice{\icmlEqualContribution} 

\begin{abstract}
Stars' chemical signatures provide invaluable insights into stellar cluster formation. This study utilized the Weisfeiler-Lehman (WL) Graph Kernel to examine a 15-dimensional elemental abundance space. Through simulating chemical distributions using normalizing flows, the effectiveness of our algorithm was affirmed. The results highlight the capability of the WL algorithm, coupled with Gaussian Process Regression, to identify patterns within elemental abundance point clouds correlated with various cluster mass functions. Notably, the WL algorithm exhibits superior interpretability, efficacy and robustness compared to deep sets and graph convolutional neural networks and enables optimal training with significantly fewer simulations ($\mathcal{O}(10)$), a reduction of at least two orders of magnitude relative to graph neural networks.
\end{abstract}

\section{Introduction}\label{sec:introduction}

Over the past decade, innovative spectroscopic surveys, such as Gaia-ESO, APOGEE, GALAH, and LAMOST, have significantly advanced our understanding of the Milky Way \citep{Gilmore2012,Luo2015,Majewski2017,Buder2020}. These surveys have generated detailed mappings of the galaxy, capturing complex chemical profiles of $10^5-10^7$ stars. Consequently, we have extensive data on up to 30 different elements within each star, surpassing the scope of past research that was limited to a few thousand high-quality stellar spectra \citep{Fuhrmann1998,Bensby2003}.

However, conventional analysis methods for this rich, high-dimensional dataset often lean on outdated models developed before the advent of comprehensive spectroscopic surveys. Many of these models oversimplify the multidimensional chemical space by focusing on chemical tracks and neglect its broader complexity \citep{Weinberg2019,Griffith2019,Griffith2021,Buck2021,Chen2022}. Furthermore, these models typically presume that stars are independently drawn from a parent distribution, failing to consider the formation of stars in clusters or aggregates.

Stars originate within clusters, spreading throughout the galaxy and leaving distinct chemical signatures \citep{Li2019,Just2023}. These tags offer precious insights into star cluster formation and dispersal \citep{Freeman2002,Spina2022}. Studying these tags affords us an understanding of the early stages of the Milky Way's evolution and its dynamic states, as the formation conditions and aggregations of stars mirror the galaxy's turbulence state \citep{Trung2021,Smith2022}.

Despite its potential, chemical tagging—the identification of unique clusters via the multidimensional space of elemental abundances—has shown limitations. While different cluster mass functions generate unique patterns within the elemental abundance space, the paucity of chemical volume currently hampers our capacity to identify individual clusters directly \citep{Ting2015,Ness2018,Price-Jones2018}. Instead, the underlying cluster mass function is inferred through robust summary statistics that capture subtle variations in point cloud morphology.
\begin{figure*}[t]
\centering
\includegraphics[width=0.8\textwidth]{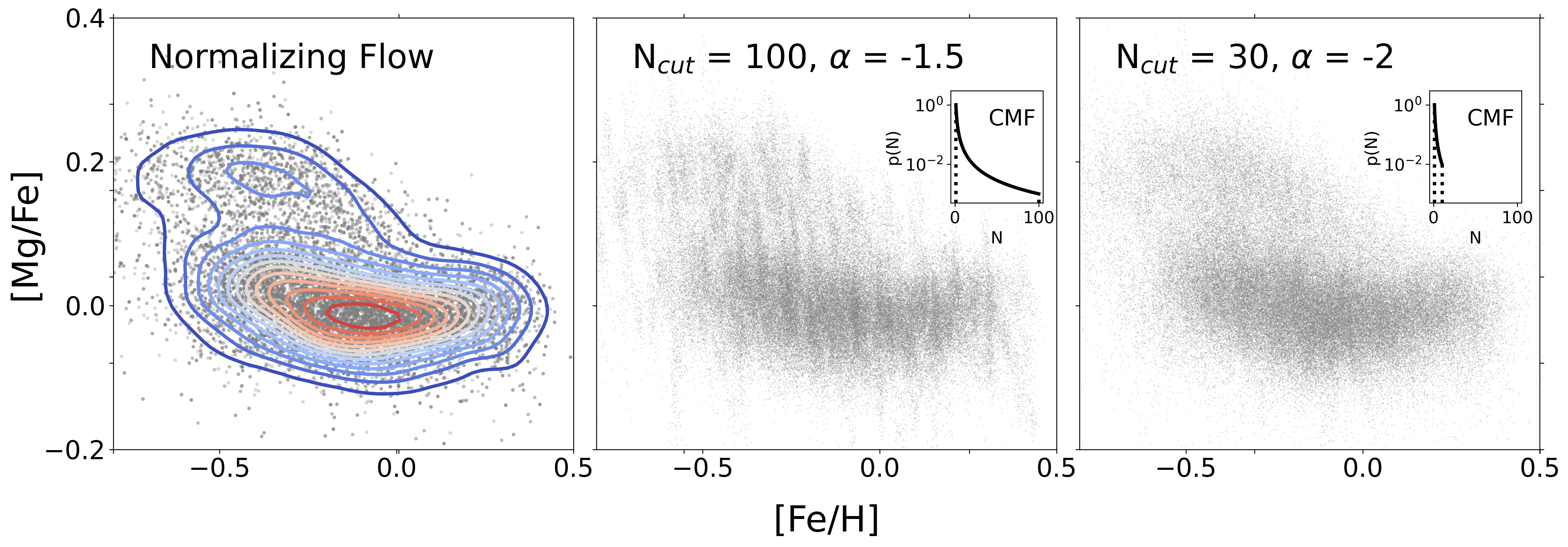}
\vskip -0.1in
\caption{Weak chemical tagging. The plot presents a 2D projection of a complex 15D elemental abundance space. The left panel displays the smooth distribution, determined from the APOGEE data via a normalizing flow. The central and right panels depict varying Milky Way stochastic chemical evolution scenarios. Incorporating larger star clusters through an elevated mass cutoff, $N_{\rm cut}$, or a subdued power-law gradient, $\alpha$, leads to increased clustering in the chemical space.}
\vskip -0.1in
\label{fig1}
\end{figure*}

This study introduces a novel approach to the complexities of statistical chemical tagging, employing a graph-based kernel method. Our methodology is inspired by the well-explored field of graph isomorphism and its ties with graph kernels, a domain extensively investigated in machine learning \citep{Kriege2020}. To our knowledge, this potent technique has not been previously utilized in astronomy. Our aim is to demonstrate the potential of the Weisfeiler-Lehman graph kernel method \citep{Shervashidze2011,Morris2017,Schulz2022}, known for its simple yet powerful formalism, in advancing the statistical study of stellar population chemical properties.

\section{Data and Simulations}
\label{sec:data}

Our model is based on a high-quality subset from the APOGEE DR16 survey \citep{Weinberg2019}, including stars from the Milky Way disk situated within a 3-13 kpc radius from the Galactic center and within 2 kpc of the Galactic plane. We focus on stars with a surface gravity between 1 and 2.5 and effective temperatures from 4100 K to 4600 K. These stars provide precise estimates of the abundance of 15 elements from APOGEE: Mg, O, Si, S, Ca, Na, Al, K, V, Cr, Mn, Fe, Co, Ni, and Cu \citep{Jonsson2020,Vincenzo2021}.

We utilize a normalizing flow model, trained on this dataset, to emulate the distribution within a 15-dimensional elemental abundance space. The model translates the complex distribution into a unit-multivariate Gaussian distribution through invertible neural networks. This allows us to sample from the distribution by first sampling from the unit Gaussian, then performing the inverse transformation. Our normalization flow includes eight units of Neural Spline Flow and GLOW \citep{Kingma2018,Durkan2019}, each containing three densely connected layers with 16 neurons.

We populate chemically analogous star clusters within this 15D space. We posit that the number of observed stars follows a power-law distribution $N \sim N^{-\alpha}$, characterized by the power law slope $\alpha$ and a high-mass cutoff of $N_{\rm cut}$ \citep{Dessauges-Zavadsky2018,Mok2019}. The elemental abundances of all stars within a cluster are derived from $\mathbb{R}^{15} \ni {\bf x} \sim \mathcal{N}({\bf x_c}, \Sigma)$, with $\Sigma$ representing the abundance dispersion due to measurement uncertainties. The centroid ${\bf x_c}$ is drawn from the smooth distribution estimated by the normalizing flow, and the measurement dispersion $\Sigma$ from the determinations from \citet{Jonsson2020}. Varying physical conditions of gas fragmentation, represented in the power law slope and high-mass cutoff, can result in subtle yet significant clustering properties of the point clouds, as demonstrated in Fig.~\ref{fig1}. 

We note that, for the high-mass cutoff $N_{\rm cut}$, it doesn't insinuate clusters initially contained such a limited number of stars. Our current million-star sample captures a minuscule fraction due to the Milky Way's vast parent sample. Notably, the current survey of $10^6$ stars generally results in a subsampling fraction of $10^{-5}$ from the parent distribution of the entire Milky Way \citep{Ting2015,Ting2016}. Consequently, $N_{\rm cut} =10$ corresponds to a star cluster of $M \simeq 10^6 M_\odot$.

\section{Methods}
\label{sec:methods}

{\bf Weisfeiler-Lehman Graph Kernel}: In the era preceding the emergence of graph neural networks (GNNs) as tools for graph representation, kernel methods were the primary choice for tasks involving graph prediction. However, one of the most prominent challenges that arose during the application of these kernel methods to graphs was formulating an efficient similarity measure. 

To address this problem, the Weisfeiler-Lehman (WL) kernel method was developed to generate features that capture the inherent structure of input graphs. For a graph $G=(V, E)$, with $V$ representing nodes and $E \subseteq V \times V$ denoting edges, the neighbourhood of a node $v$ is defined as $\mathcal{N}(v) = \{u \mid (u,v) \in E\}$. In our situation, we start with point clouds in the elemental abundance space that lack edges. We then impose a graph structure by connecting nodes that are within a Euclidean distance of less than 0.15 dex in a 15-dimensional space. The radius for this is obtained through a hyperparameter grid search, and the degree of individual nodes serves as the initial node attributes.

\begin{figure}[t]
\centering
\includegraphics[width=0.48\textwidth]{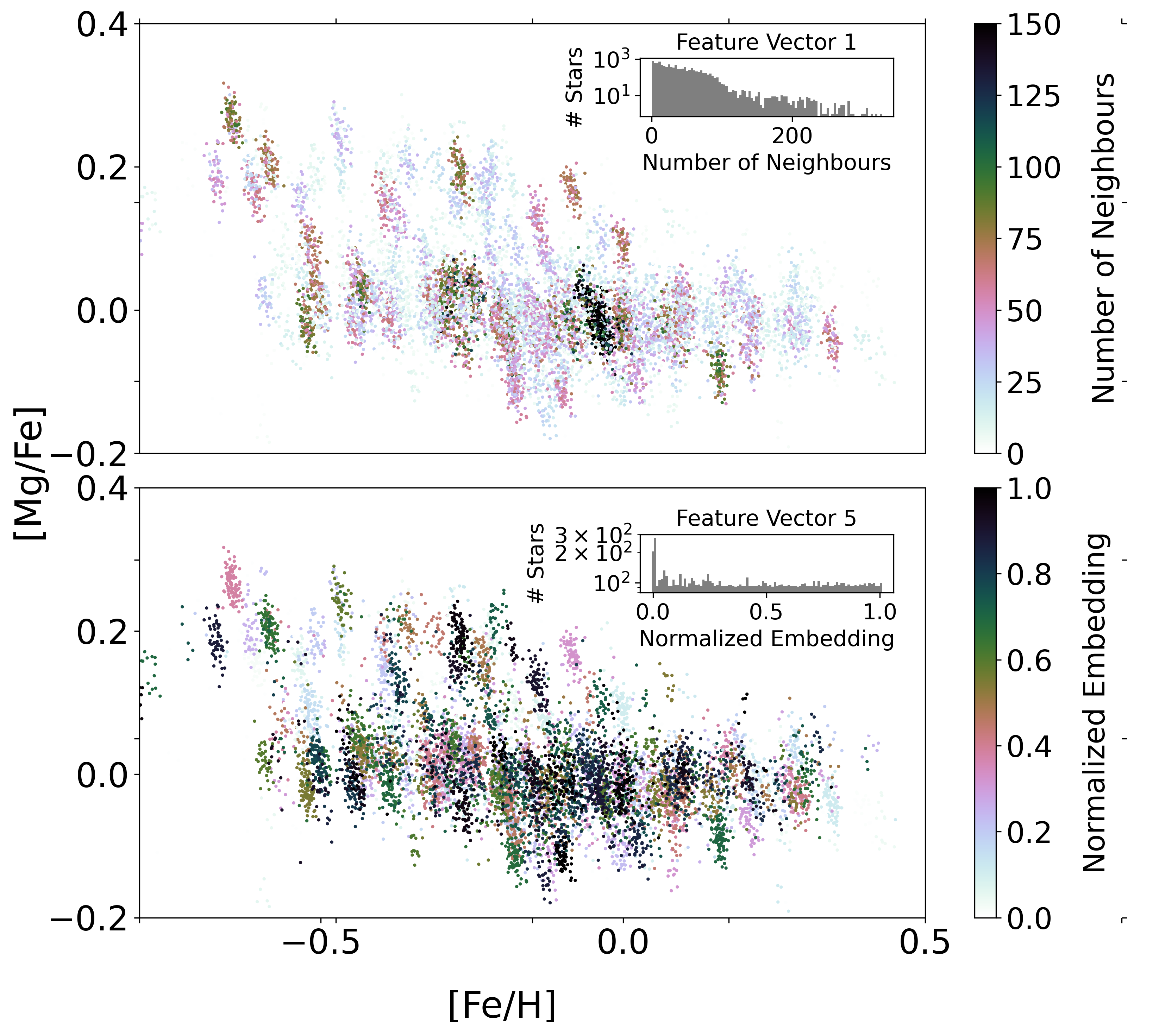}
\vskip -0.1in
\caption{Initiating with the neighbour count within a predetermined radius (Top Panel), the WL kernel creates an aggregate embedding tuple for hierarchical similarity analysis. Post five iterations, nodes are colour-coded per WL embedding (Bottom Panel). The similarity between graphs is gauged via the dot product of the histogram of these successive embeddings, functioning as feature vectors (shown in the Inset Panels).}
\vskip -0.2in
\label{fig2}
\end{figure}

The WL algorithm is designed to refine node attributes iteratively by incorporating local information. For a given node $v$, a multiset of neighbouring attributes is established and then distilled into a unique embedding $l_i$ via an injective function $f$. The WL kernel method determines the graph similarity by conducting a set number of iterations to create a feature map of the graph, $\phi(G)$, which is the histogram vector of the node embedding $l_i$. The pseudocodes are listed in Table~\ref{alg:wl} (refer to Fig.\ref{fig2}). The similarity of two graphs, or the kernel $k(G, G')$, is then defined as the inner product of their respective features: $k(G, G') = \phi(G)^T \phi(G')$. The optimal number of iterations, found to be five in our case, through hyperparameter grid search.

We perform inference on the target label ${\bf y} = (\boldsymbol{\alpha}, \boldsymbol{N_{\rm cut}})$ using a zero-mean Gaussian Process (GP) and the graph kernel as a similarity measure for observed graphs. For any new graph $G_*$, its target value $y_*$ is predicted by a weighted sum of known ${\bf y}$ values, with weights derived from the WL kernel-based similarity between $G_*$ and the training graphs ${\bf G}$. The mean prediction of $y_*$ from GP is computed as $y_* = k(G_*,{\bf G}) k({\bf G},{\bf G})^{-1} \boldsymbol{y}$, indicating that the new graph's target value prediction depends on its similarity to training graphs and the training labels

To holistically evaluate our method, we juxtapose it with two techniques: Deep Sets and Graph Neural Networks.

{\bf Deep Set Method}: In this model, feature vectors $\tau(u) \in \mathbb{R}^{k+n}$ are constructed for each node $u$ in a given graph realization $G$. This includes quantifying neighbours within a series of $k$ radii, $r_k \in \{0.01, 0.05, 0.10, 0.15, 0.20\}$, and using $n$-dimensional positional information to distinguish isolated and densely populated clusters. The model output on an elemental abundance graph $G$ is then given by, $\phi \left( \bigoplus_{u \in G} \psi(\tau(u)) \right)$. Here, $\psi$ transforms the feature vectors and is given by a learnable multi-layer perceptron (MLP) with a single layer, $\bigoplus$ is a mean aggregation operation, and $\phi$ is another single-layer MLP that maps the aggregated vector representation to parameters $y = (\alpha, N_{\rm cut})$.

\begin{algorithm}[tb]
   \caption{The Weisfeiler-Lehman Algorithm}
   \label{alg:wl}
\begin{algorithmic}
   \STATE {\bfseries Input:} Graphs $G_1,\ldots,G_N$, Total number of WL iterations $h$, Injective function $f$
   \FOR{$i=1$ {\bfseries to} $N$}
   \STATE $l_i(v) \leftarrow |\mathcal{N}(v)|, \forall v \in G_i$
   \FOR{$k=1$ to $h$}
   \STATE $l_i'(v) \leftarrow f(l_i(v), \{\!\!\{ l_i(u) \mid u \in \mathcal{N}(v)\}\!\!\}), \forall v \in G_i$
   \STATE $l_i \leftarrow l_i'$
   \ENDFOR
   \ENDFOR
   \STATE {\bfseries Return} Histogram of graph embeddings $l_i:V_i\rightarrow \mathbb{N}$ , from all iterations as features for $G_1,\ldots, G_N$.
\end{algorithmic}
\end{algorithm}

    
       

{\bf Convolutional Graph Neural Network (ConvGNN)}: The ConvGNN model addresses non-uniform data distributions by preserving isolated clusters' statistics through local aggregation. Each node $u \in G$ has a hidden representation $\mathbf{h}_u$, updated through a single GNN message-passing layer, and defined by $\mathbf{h}_u = \phi\left( \tau(u), \bigoplus_{v \in \mathcal{N}(u)} \psi(\tau(v)) \right)$, where $\psi$ and $\phi$ are learnable MLPs and $\bigoplus$ is a mean aggregation function. The node-level representations are averaged with a readout function $\mathcal{R}$ and passed through another single-layer MLP $\rho$, $\rho\left(\mathcal{R}(\{\mathbf{h}_u \mid u \in G\}\right)$ to infer the physical parameter $y = (\alpha, N_{\rm cut})$, effectively leveraging the structural information of the graphs.


\section{Results}
\label{sec:results}

This study demonstrates the potent potential of the WL algorithm in mining intrinsic physical parameters from elemental abundance point clouds. Through $5,000$ distinct chemical abundance simulations—each characterized by a unique power-law slope $\alpha$ and high-mass cutoff $N_{\rm cut}$—we evaluate the performance and versatility of the WL algorithm.

\begin{figure}[t]
\centering
\includegraphics[width=0.5\textwidth]{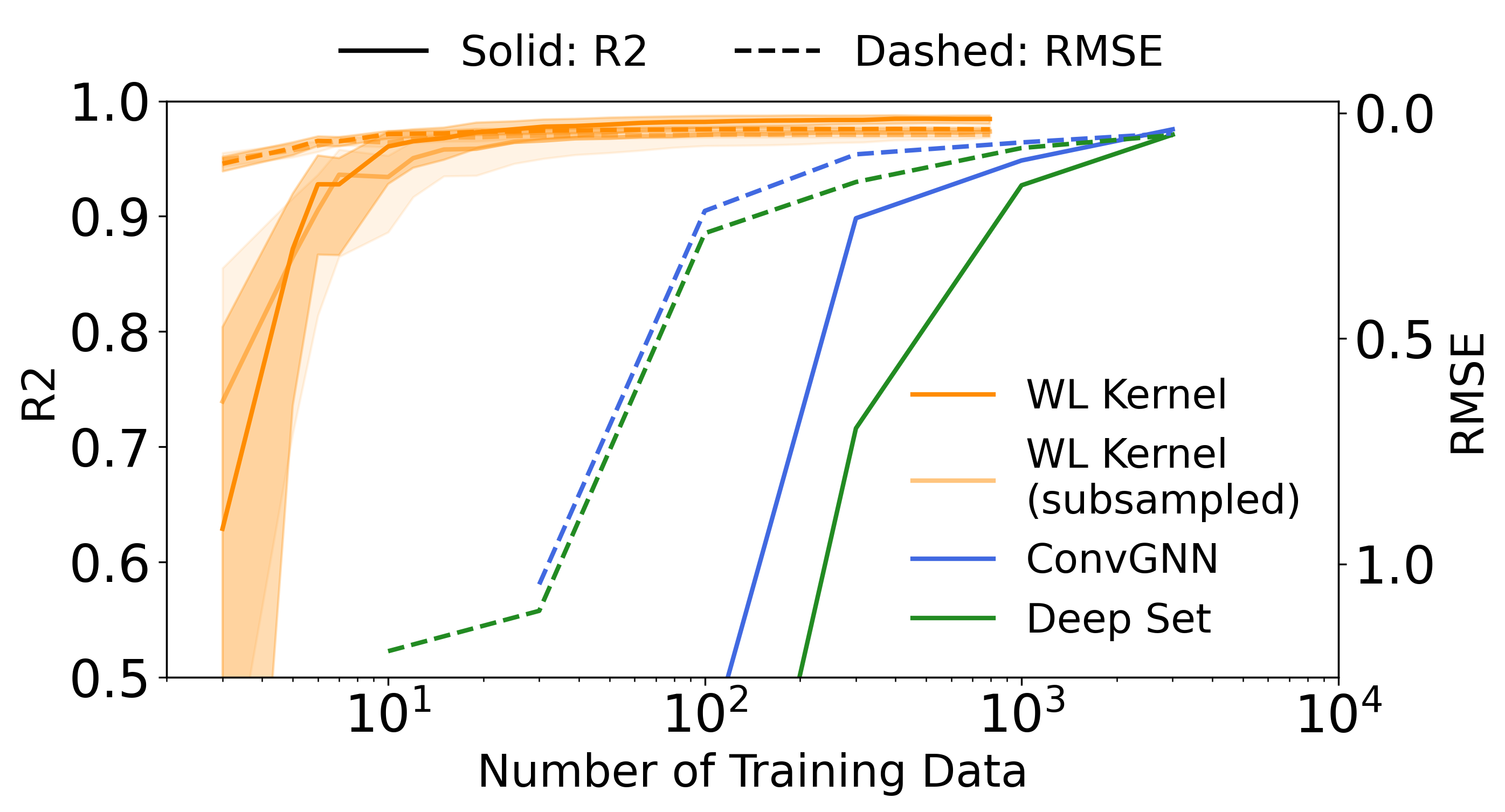}
\vskip -0.1in
\caption{Performance of the WL kernel in weak chemical tagging. The plot presents the R² score and Root Mean Squared Error (RMSE) as a function of training data size, underscoring the WL kernel's effectiveness in predicting complex data structures as opposed to Deep Sets and ConvGNNs.}
\vskip -0.2in
\label{fig3}
\end{figure}

{\bf Efficacy and Robustness:} We measure efficacy by the minimum training samples necessary to ensure reliable results, quantified via the R² score and RMSE of the power-law slope recovery $\alpha$ (refer to Fig~\ref{fig3}). Considering the extensive training times of ConvGNN, we narrow our scenario by setting $N_{\rm cut}$ at 100 and taking $\alpha \sim \mathcal{U}(-2.5,-1.5)$.

The WL algorithm's edge over deep-set and ConvGNN methods is unmistakable from Fig~\ref{fig3}, excelling in interpreting nuanced elemental abundance data. Even with minimal simulations—on the order of $\mathcal{O}(1)-\mathcal{O}(10)$—the WL kernel accurately infers the underlying parameters, outperforming alternatives that rely on significantly larger datasets—on the order of $\mathcal{O}(1000)$. This proficiency is particularly advantageous when applying the WL algorithm to realistic hydrodynamical simulations, where the training limitations of ConvGNN would prove problematic.

{\bf Scalability and Interpretability:} The impressive performance of the WL algorithm underlines the importance of a strong inductive bias in assessing the subtleties of star clustering in chemical space. Fig.~\ref{fig2} showcases how the WL approach, which implements a predetermined message passing and aggregation mechanism, as opposed to ConvGNNs, leads to robust results and enhanced interpretability. Hierarchical embedding of each node, for example, aids in identifying systematic biases such as potential biasing signals from open clusters or globular clusters.

Additionally, recognizing hierarchical signals and the inherent inductive bias enables us to make practical simplifications without substantial accuracy compromise. As demonstrated by Fig~\ref{fig3}, even when only one in every five nodes is randomly selected in the graph during message passing, the robustness of the results is preserved. This is mainly attributable to the similarity in the node embedding in neighbouring features. Such a reduction, cutting computation by a factor of $5^2=25$, barely impacts the overall outcome

\begin{figure}[t]
    \centering
     \includegraphics[width=0.45\textwidth]{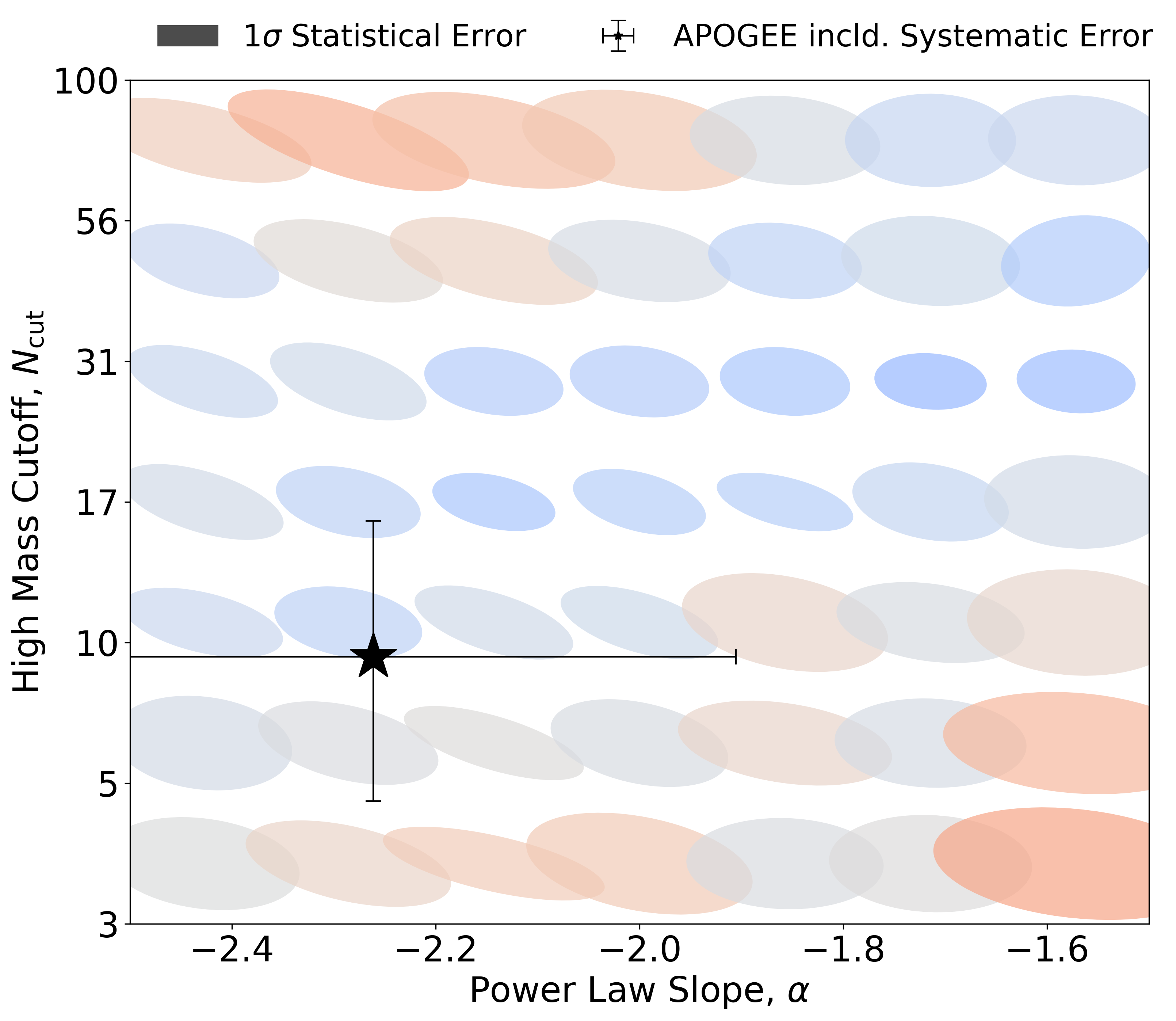}
     \vskip -0.1in
     \caption{Recovery of the power-law slope and high-mass cutoff via WL kernel. The ellipses illustrate the statistical errors for point clouds of $2 \times 10^4$ stars measured empirically from the mock tests. The symbol represents the limit when applying the algorithm to a well-curated, albeit incomplete, APOGEE sample of $2 \times 10^4$ stars.}
     \vskip -0.2in
     \label{fig4}
\end{figure} 

{\bf Prospects for Weak Chemical Tagging:} Exploring the practical utility of the WL kernel, we turn our attention to Fig~\ref{fig4}. We adjust both the high-mass cutoff $N_{\rm cut}$ and the power-law slope $\alpha$. For this final implementation, we use a training dataset of 500. The ellipses in Fig~\ref{fig4} depict the empirical uncertainty derived from the test dataset, assessed via 10-fold cross-validation using the $5,000$ simulations. While some extreme conditions lead to marginal uncertainty degradation, the WL algorithm consistently produces a suitable statistical uncertainty range across a broad parameter space at the level of $0.1-0.2$ dex in $\alpha$ and $0.1-0.15$ dex in the logarithmic of $N_{\rm cut}$ with $2 \times 10^4$ stars.

For an initial trial, we apply the high-quality APOGEE sample of $2 \times 10^4$ stars, yielding optimal values of $\alpha = -2.26^{+0.36}_{-0.32}$ and $N_{\rm cut} = 9.4^{+7.0}_{-4.2}$, with uncertainties calculated through bootstrapping. This approach accounts for both statistical and systematic errors due to the model assumption. The derived power-law slope, $\alpha$, aligns with young clusters' research \citep{Dessauges-Zavadsky2018}. For the high-mass cutoff, the $N_{\rm cut}$ equals $9.4$ for a sample size of $2 \times 10^4$, which when factored with the sampling rate, corresponds to $10^{7}-10^{8} {\rm M}_\odot$ (see Section~\ref{sec:data}), consistent with the globular cluster masses. The anticipated full paper will probe the global CMF constraint with a larger sample ($10^7$ stars) from various spectroscopic surveys, and further investigate its evolution by delving deeper into the elemental abundance space in tandem with stellar ages.

\section{Broader Impact}

The structure of the universe manifests in complex patterns in our observations. Decoding these patterns—astrostatistics' primary challenge—uncovers the physical parameters governing the system. Despite its power, astronomical analysis often neglects the rich information embedded in hierarchical structures, typically focusing on scalar and vector fields. Graph neural networks have disrupted this traditional field-only analysis, leveraging raw observational data in the form of point clouds of features like positions, velocities, and elemental abundances. However, they grapple with interpretability and training convergence.

This study puts forth graph kernels that can rival graph neural networks. The WL kernel's success largely stems from its strong inductive bias, effectively harnessing hierarchical information from astronomical point clouds. By addressing the challenge of weak chemical tagging, we introduce a first-principle approach with the potential to further Galactic Archaeology studies. Furthermore, our techniques could inspire new graph neural network applications, fostering a more statistical and fundamental approach.

\bibliography{example_paper}

\begin{thebibliography}{31}
\providecommand{\natexlab}[1]{#1}
\providecommand{\url}[1]{\texttt{#1}}
\expandafter\ifx\csname urlstyle\endcsname\relax
  \providecommand{\doi}[1]{doi: #1}\else
  \providecommand{\doi}{doi: \begingroup \urlstyle{rm}\Url}\fi

\bibitem[{Bensby} et~al.(2003){Bensby}, {Feltzing}, and
  {Lundstr{\"o}m}]{Bensby2003}
{Bensby}, T., {Feltzing}, S., and {Lundstr{\"o}m}, I.
\newblock {Elemental abundance trends in the Galactic thin and thick disks as
  traced by nearby F and G dwarf stars}.
\newblock \emph{\aap}, 410:\penalty0 527--551, November 2003.
\newblock \doi{10.1051/0004-6361:20031213}.

\bibitem[{Buck} et~al.(2021){Buck}, {Rybizki}, {Buder}, {Obreja}, {Macci{\`o}},
  {Pfrommer}, {Steinmetz}, and {Ness}]{Buck2021}
{Buck}, T., {Rybizki}, J., {Buder}, S., {Obreja}, A., {Macci{\`o}}, A.~V.,
  {Pfrommer}, C., {Steinmetz}, M., and {Ness}, M.
\newblock {The challenge of simultaneously matching the observed diversity of
  chemical abundance patterns in cosmological hydrodynamical simulations}.
\newblock \emph{\mnras}, 508\penalty0 (3):\penalty0 3365--3387, December 2021.
\newblock \doi{10.1093/mnras/stab2736}.

\bibitem[{Buder} et~al.(2020){Buder}, {Sharma}, {Kos}, {Amarsi}, {Nordlander},
  {Lind}, {Martell}, {Asplund}, {Bland-Hawthorn}, {Casey}, {De Silva},
  {D'Orazi}, {Freeman}, {Hayden}, {Lewis}, {Lin}, {Schlesinger}, {Simpson},
  {Stello}, {Zucker}, {Zwitter}, {Beeson}, {Buck}, {Casagrande}, {Clark},
  {Cotar}, {Da Costa}, {de Grijs}, {Feuillet}, {Horner}, {Khanna}, {Kafle},
  {Liu}, {Montet}, {Nandakumar}, {Nataf}, {Ness}, {Spina}, {Traven},
  {Tepper-Garcia}, {Ting}, {Vogrincic}, {Wittenmyer}, {Zerjal}, and {the GALAH
  collaboration}]{Buder2020}
{Buder}, S., {Sharma}, S., {Kos}, J., {Amarsi}, A.~M., {Nordlander}, T.,
  {Lind}, K., {Martell}, S.~L., {Asplund}, M., {Bland-Hawthorn}, J., {Casey},
  A.~R., {De Silva}, G.~M., {D'Orazi}, V., {Freeman}, K.~C., {Hayden}, M.~R.,
  {Lewis}, G.~F., {Lin}, J., {Schlesinger}, K.~J., {Simpson}, J.~D., {Stello},
  D., {Zucker}, D.~B., {Zwitter}, T., {Beeson}, K.~L., {Buck}, T.,
  {Casagrande}, L., {Clark}, J.~T., {Cotar}, K., {Da Costa}, G.~S., {de Grijs},
  R., {Feuillet}, D., {Horner}, J., {Khanna}, S., {Kafle}, P.~R., {Liu}, F.,
  {Montet}, B.~T., {Nandakumar}, G., {Nataf}, D.~M., {Ness}, M.~K., {Spina},
  L., {Traven}, G., {Tepper-Garcia}, T., {Ting}, Y.-S., {Vogrincic}, R.,
  {Wittenmyer}, R.~A., {Zerjal}, M., and {the GALAH collaboration}.
\newblock {The GALAH+ Survey: Third Data Release}.
\newblock \emph{arXiv e-prints}, art. arXiv:2011.02505, November 2020.

\bibitem[{Chen} et~al.(2022){Chen}, {Hayden}, {Sharma}, {Bland-Hawthorn},
  {Kobayashi}, and {Karakas}]{Chen2022}
{Chen}, B., {Hayden}, M.~R., {Sharma}, S., {Bland-Hawthorn}, J., {Kobayashi},
  C., and {Karakas}, A.~I.
\newblock {Chemical Evolution with Radial Mixing Redux: Extending beyond the
  Solar Neighborhood}.
\newblock \emph{arXiv e-prints}, art. arXiv:2204.11413, April 2022.
\newblock \doi{10.48550/arXiv.2204.11413}.

\bibitem[{Dessauges-Zavadsky} \& {Adamo}(2018){Dessauges-Zavadsky} and
  {Adamo}]{Dessauges-Zavadsky2018}
{Dessauges-Zavadsky}, M. and {Adamo}, A.
\newblock {First constraints on the stellar mass function of star-forming
  clumps at the peak of cosmic star formation}.
\newblock \emph{\mnras}, 479\penalty0 (1):\penalty0 L118--L122, September 2018.
\newblock \doi{10.1093/mnrasl/sly112}.

\bibitem[{Durkan} et~al.(2019){Durkan}, {Bekasov}, {Murray}, and
  {Papamakarios}]{Durkan2019}
{Durkan}, C., {Bekasov}, A., {Murray}, I., and {Papamakarios}, G.
\newblock {Neural Spline Flows}.
\newblock \emph{arXiv e-prints}, art. arXiv:1906.04032, June 2019.

\bibitem[{Freeman} \& {Bland-Hawthorn}(2002){Freeman} and
  {Bland-Hawthorn}]{Freeman2002}
{Freeman}, K. and {Bland-Hawthorn}, J.
\newblock {The New Galaxy: Signatures of Its Formation}.
\newblock \emph{\araa}, 40:\penalty0 487--537, January 2002.
\newblock \doi{10.1146/annurev.astro.40.060401.093840}.

\bibitem[{Fuhrmann}(1998)]{Fuhrmann1998}
{Fuhrmann}, K.
\newblock {Nearby stars of the Galactic disk and halo}.
\newblock \emph{\aap}, 338:\penalty0 161--183, October 1998.

\bibitem[{Gilmore} et~al.(2012){Gilmore}, {Randich}, {Asplund}, {Binney},
  {Bonifacio}, {Drew}, {Feltzing}, {Ferguson}, {Jeffries}, {Micela},
  {Negueruela}, {Prusti}, {Rix}, {Vallenari}, {Alfaro}, {Allende-Prieto},
  {Babusiaux}, {Bensby}, {Blomme}, {Bragaglia}, {Flaccomio}, {Fran{\c{c}}ois},
  {Irwin}, {Koposov}, {Korn}, {Lanzafame}, {Pancino}, {Paunzen},
  {Recio-Blanco}, {Sacco}, {Smiljanic}, {Van Eck}, {Walton}, {Aden}, {Aerts},
  {Affer}, {Alcala}, {Altavilla}, {Alves}, {Antoja}, {Arenou}, {Argiroffi},
  {Asensio Ramos}, {Bailer-Jones}, {Balaguer-Nunez}, {Bayo}, {Barbuy},
  {Barisevicius}, {Barrado y Navascues}, {Battistini}, {Bellas Velidis},
  {Bellazzini}, {Belokurov}, {Bergemann}, {Bertelli}, {Biazzo}, {Bienayme},
  {Bland-Hawthorn}, {Boeche}, {Bonito}, {Boudreault}, {Bouvier}, {Brandao},
  {Brown}, {de Bruijne}, {Burleigh}, {Caballero}, {Caffau}, {Calura},
  {Capuzzo-Dolcetta}, {Caramazza}, {Carraro}, {Casagrande}, {Casewell},
  {Chapman}, {Chiappini}, {Chorniy}, {Christlieb}, {Cignoni}, {Cocozza},
  {Colless}, {Collet}, {Collins}, {Correnti}, {Covino}, {Crnojevic}, {Cropper},
  {Cunha}, {Damiani}, {David}, {Delgado}, {Duffau}, {Edvardsson}, {Eldridge},
  {Enke}, {Eriksson}, {Evans}, {Eyer}, {Famaey}, {Fellhauer}, {Ferreras},
  {Figueras}, {Fiorentino}, {Flynn}, {Folha}, {Franciosini}, {Frasca},
  {Freeman}, {Fremat}, {Friel}, {Gaensicke}, {Gameiro}, {Garzon}, {Geier},
  {Geisler}, {Gerhard}, {Gibson}, {Gomboc}, {Gomez}, {Gonzalez-Fernandez},
  {Gonzalez Hernandez}, {Gosset}, {Grebel}, {Greimel}, {Groenewegen},
  {Grundahl}, {Guarcello}, {Gustafsson}, {Hadrava}, {Hatzidimitriou}, {Hambly},
  {Hammersley}, {Hansen}, {Haywood}, {Heber}, {Heiter}, {Held}, {Helmi},
  {Hensler}, {Herrero}, {Hill}, {Hodgkin}, {Huelamo}, {Huxor}, {Ibata},
  {Jackson}, {de Jong}, {Jonker}, {Jordan}, {Jordi}, {Jorissen}, {Katz},
  {Kawata}, {Keller}, {Kharchenko}, {Klement}, {Klutsch}, {Knude}, {Koch},
  {Kochukhov}, {Kontizas}, {Koubsky}, {Lallement}, {de Laverny}, {van Leeuwen},
  {Lemasle}, {Lewis}, {Lind}, {Lindstrom}, {Lobel}, {Lopez Santiago}, {Lucas},
  {Ludwig}, {Lueftinger}, {Magrini}, {Maiz Apellaniz}, {Maldonado}, {Marconi},
  {Marino}, {Martayan}, {Martinez-Valpuesta}, {Matijevic}, {McMahon},
  {Messina}, {Meyer}, {Miglio}, {Mikolaitis}, {Minchev}, {Minniti}, {Moitinho},
  {Momany}, {Monaco}, {Montalto}, {Monteiro}, {Monier}, {Montes}, {Mora},
  {Moraux}, {Morel}, {Mowlavi}, {Mucciarelli}, {Munari}, {Napiwotzki},
  {Nardetto}, {Naylor}, {Naze}, {Nelemans}, {Okamoto}, {Ortolani}, {Pace},
  {Palla}, {Palous}, {Parker}, {Penarrubia}, {Pillitteri}, {Piotto}, {Posbic},
  {Prisinzano}, {Puzeras}, {Quirrenbach}, {Ragaini}, {Read}, {Read}, {Reyle},
  {De Ridder}, {Robichon}, {Robin}, {Roeser}, {Romano}, {Royer}, {Ruchti},
  {Ruzicka}, {Ryan}, {Ryde}, {Santos}, {Sanz Forcada}, {Sarro Baro},
  {Sbordone}, {Schilbach}, {Schmeja}, {Schnurr}, {Schoenrich}, {Scholz},
  {Seabroke}, {Sharma}, {De Silva}, {Smith}, {Solano}, {Sordo}, {Soubiran},
  {Sousa}, {Spagna}, {Steffen}, {Steinmetz}, {Stelzer}, {Stempels},
  {Tabernero}, {Tautvaisiene}, {Thevenin}, {Torra}, {Tosi}, {Tolstoy}, {Turon},
  {Walker}, {Wambsganss}, {Worley}, {Venn}, {Vink}, {Wyse}, {Zaggia},
  {Zeilinger}, {Zoccali}, {Zorec}, {Zucker}, {Zwitter}, and {Gaia-ESO Survey
  Team}]{Gilmore2012}
{Gilmore}, G., {Randich}, S., {Asplund}, M., {Binney}, J., {Bonifacio}, P.,
  {Drew}, J., {Feltzing}, S., {Ferguson}, A., {Jeffries}, R., {Micela}, G.,
  {Negueruela}, I., {Prusti}, T., {Rix}, H.~W., {Vallenari}, A., {Alfaro}, E.,
  {Allende-Prieto}, C., {Babusiaux}, C., {Bensby}, T., {Blomme}, R.,
  {Bragaglia}, A., {Flaccomio}, E., {Fran{\c{c}}ois}, P., {Irwin}, M.,
  {Koposov}, S., {Korn}, A., {Lanzafame}, A., {Pancino}, E., {Paunzen}, E.,
  {Recio-Blanco}, A., {Sacco}, G., {Smiljanic}, R., {Van Eck}, S., {Walton},
  N., {Aden}, D., {Aerts}, C., {Affer}, L., {Alcala}, J.~M., {Altavilla}, G.,
  {Alves}, J., {Antoja}, T., {Arenou}, F., {Argiroffi}, C., {Asensio Ramos},
  A., {Bailer-Jones}, C., {Balaguer-Nunez}, L., {Bayo}, A., {Barbuy}, B.,
  {Barisevicius}, G., {Barrado y Navascues}, D., {Battistini}, C., {Bellas
  Velidis}, I., {Bellazzini}, M., {Belokurov}, V., {Bergemann}, M., {Bertelli},
  G., {Biazzo}, K., {Bienayme}, O., {Bland-Hawthorn}, J., {Boeche}, C.,
  {Bonito}, S., {Boudreault}, S., {Bouvier}, J., {Brandao}, I., {Brown}, A.,
  {de Bruijne}, J., {Burleigh}, M., {Caballero}, J., {Caffau}, E., {Calura},
  F., {Capuzzo-Dolcetta}, R., {Caramazza}, M., {Carraro}, G., {Casagrande}, L.,
  {Casewell}, S., {Chapman}, S., {Chiappini}, C., {Chorniy}, Y., {Christlieb},
  N., {Cignoni}, M., {Cocozza}, G., {Colless}, M., {Collet}, R., {Collins}, M.,
  {Correnti}, M., {Covino}, E., {Crnojevic}, D., {Cropper}, M., {Cunha}, M.,
  {Damiani}, F., {David}, M., {Delgado}, A., {Duffau}, S., {Edvardsson}, B.,
  {Eldridge}, J., {Enke}, H., {Eriksson}, K., {Evans}, N.~W., {Eyer}, L.,
  {Famaey}, B., {Fellhauer}, M., {Ferreras}, I., {Figueras}, F., {Fiorentino},
  G., {Flynn}, C., {Folha}, D., {Franciosini}, E., {Frasca}, A., {Freeman}, K.,
  {Fremat}, Y., {Friel}, E., {Gaensicke}, B., {Gameiro}, J., {Garzon}, F.,
  {Geier}, S., {Geisler}, D., {Gerhard}, O., {Gibson}, B., {Gomboc}, A.,
  {Gomez}, A., {Gonzalez-Fernandez}, C., {Gonzalez Hernandez}, J., {Gosset},
  E., {Grebel}, E., {Greimel}, R., {Groenewegen}, M., {Grundahl}, F.,
  {Guarcello}, M., {Gustafsson}, B., {Hadrava}, P., {Hatzidimitriou}, D.,
  {Hambly}, N., {Hammersley}, P., {Hansen}, C., {Haywood}, M., {Heber}, U.,
  {Heiter}, U., {Held}, E., {Helmi}, A., {Hensler}, G., {Herrero}, A., {Hill},
  V., {Hodgkin}, S., {Huelamo}, N., {Huxor}, A., {Ibata}, R., {Jackson}, R.,
  {de Jong}, R., {Jonker}, P., {Jordan}, S., {Jordi}, C., {Jorissen}, A.,
  {Katz}, D., {Kawata}, D., {Keller}, S., {Kharchenko}, N., {Klement}, R.,
  {Klutsch}, A., {Knude}, J., {Koch}, A., {Kochukhov}, O., {Kontizas}, M.,
  {Koubsky}, P., {Lallement}, R., {de Laverny}, P., {van Leeuwen}, F.,
  {Lemasle}, B., {Lewis}, G., {Lind}, K., {Lindstrom}, H.~P.~E., {Lobel}, A.,
  {Lopez Santiago}, J., {Lucas}, P., {Ludwig}, H., {Lueftinger}, T., {Magrini},
  L., {Maiz Apellaniz}, J., {Maldonado}, J., {Marconi}, G., {Marino}, A.,
  {Martayan}, C., {Martinez-Valpuesta}, I., {Matijevic}, G., {McMahon}, R.,
  {Messina}, S., {Meyer}, M., {Miglio}, A., {Mikolaitis}, S., {Minchev}, I.,
  {Minniti}, D., {Moitinho}, A., {Momany}, Y., {Monaco}, L., {Montalto}, M.,
  {Monteiro}, M.~J., {Monier}, R., {Montes}, D., {Mora}, A., {Moraux}, E.,
  {Morel}, T., {Mowlavi}, N., {Mucciarelli}, A., {Munari}, U., {Napiwotzki},
  R., {Nardetto}, N., {Naylor}, T., {Naze}, Y., {Nelemans}, G., {Okamoto}, S.,
  {Ortolani}, S., {Pace}, G., {Palla}, F., {Palous}, J., {Parker}, R.,
  {Penarrubia}, J., {Pillitteri}, I., {Piotto}, G., {Posbic}, H., {Prisinzano},
  L., {Puzeras}, E., {Quirrenbach}, A., {Ragaini}, S., {Read}, J., {Read}, M.,
  {Reyle}, C., {De Ridder}, J., {Robichon}, N., {Robin}, A., {Roeser}, S.,
  {Romano}, D., {Royer}, F., {Ruchti}, G., {Ruzicka}, A., {Ryan}, S., {Ryde},
  N., {Santos}, N., {Sanz Forcada}, J., {Sarro Baro}, L.~M., {Sbordone}, L.,
  {Schilbach}, E., {Schmeja}, S., {Schnurr}, O., {Schoenrich}, R., {Scholz},
  R.~D., {Seabroke}, G., {Sharma}, S., {De Silva}, G., {Smith}, M., {Solano},
  E., {Sordo}, R., {Soubiran}, C., {Sousa}, S., {Spagna}, A., {Steffen}, M.,
  {Steinmetz}, M., {Stelzer}, B., {Stempels}, E., {Tabernero}, H.,
  {Tautvaisiene}, G., {Thevenin}, F., {Torra}, J., {Tosi}, M., {Tolstoy}, E.,
  {Turon}, C., {Walker}, M., {Wambsganss}, J., {Worley}, C., {Venn}, K.,
  {Vink}, J., {Wyse}, R., {Zaggia}, S., {Zeilinger}, W., {Zoccali}, M.,
  {Zorec}, J., {Zucker}, D., {Zwitter}, T., and {Gaia-ESO Survey Team}.
\newblock {The Gaia-ESO Public Spectroscopic Survey}.
\newblock \emph{The Messenger}, 147:\penalty0 25--31, March 2012.

\bibitem[{Griffith} et~al.(2019){Griffith}, {Johnson}, and
  {Weinberg}]{Griffith2019}
{Griffith}, E., {Johnson}, J.~A., and {Weinberg}, D.~H.
\newblock {Abundance Ratios in GALAH DR2 and Their Implications for
  Nucleosynthesis}.
\newblock \emph{\apj}, 886\penalty0 (2):\penalty0 84, December 2019.
\newblock \doi{10.3847/1538-4357/ab4b5d}.

\bibitem[{Griffith} et~al.(2020){Griffith}, {Weinberg}, {Johnson}, {Beaton},
  {Garc{\'\i}a-Hern{\'a}ndez}, {Hasselquist}, {Holtzman}, {Johnson},
  {J{\"o}nsson}, {Lane}, {Nataf}, and {Roman-Lopes}]{Griffith2021}
{Griffith}, E., {Weinberg}, D.~H., {Johnson}, J.~A., {Beaton}, R.,
  {Garc{\'\i}a-Hern{\'a}ndez}, D.~A., {Hasselquist}, S., {Holtzman}, J.,
  {Johnson}, J.~W., {J{\"o}nsson}, H., {Lane}, R.~R., {Nataf}, D.~M., and
  {Roman-Lopes}, A.
\newblock {The Similarity of Abundance Ratio Trends and Nucleosynthetic
  Patterns in the Milky Way Disk and Bulge}.
\newblock \emph{arXiv e-prints}, art. arXiv:2009.05063, September 2020.

\bibitem[{Ha} et~al.(2021){Ha}, {Li}, {Xu}, {Kounkel}, and {Li}]{Trung2021}
{Ha}, T., {Li}, Y., {Xu}, S., {Kounkel}, M., and {Li}, H.
\newblock {Measuring Turbulence with Young Stars in the Orion Complex}.
\newblock \emph{\apjl}, 907\penalty0 (2):\penalty0 L40, February 2021.
\newblock \doi{10.3847/2041-8213/abd8c9}.

\bibitem[{J{\"o}nsson} et~al.(2020){J{\"o}nsson}, {Holtzman}, {Allende Prieto},
  {Cunha}, {Garc{\'\i}a-Hern{\'a}ndez}, {Hasselquist}, {Masseron}, {Osorio},
  {Shetrone}, {Smith}, {Stringfellow}, {Bizyaev}, {Edvardsson}, {Majewski},
  {M{\'e}sz{\'a}ros}, {Souto}, {Zamora}, {Beaton}, {Bovy}, {Donor},
  {Pinsonneault}, {Poovelil}, and {Sobeck}]{Jonsson2020}
{J{\"o}nsson}, H., {Holtzman}, J.~A., {Allende Prieto}, C., {Cunha}, K.,
  {Garc{\'\i}a-Hern{\'a}ndez}, D.~A., {Hasselquist}, S., {Masseron}, T.,
  {Osorio}, Y., {Shetrone}, M., {Smith}, V., {Stringfellow}, G.~S., {Bizyaev},
  D., {Edvardsson}, B., {Majewski}, S.~R., {M{\'e}sz{\'a}ros}, S., {Souto}, D.,
  {Zamora}, O., {Beaton}, R.~L., {Bovy}, J., {Donor}, J., {Pinsonneault},
  M.~H., {Poovelil}, V.~J., and {Sobeck}, J.
\newblock {APOGEE Data and Spectral Analysis from SDSS Data Release 16: Seven
  Years of Observations Including First Results from APOGEE-South}.
\newblock \emph{\aj}, 160\penalty0 (3):\penalty0 120, September 2020.
\newblock \doi{10.3847/1538-3881/aba592}.

\bibitem[{Just} et~al.(2023){Just}, {Piskunov}, {Klos}, {Kovaleva}, and
  {Polyachenko}]{Just2023}
{Just}, A., {Piskunov}, A.~E., {Klos}, J.~H., {Kovaleva}, D.~A., and
  {Polyachenko}, E.~V.
\newblock {Global survey of star clusters in the Milky Way. VII. Tidal
  parameters and mass function}.
\newblock \emph{\aap}, 672:\penalty0 A187, April 2023.
\newblock \doi{10.1051/0004-6361/202244723}.

\bibitem[{Kingma} \& {Dhariwal}(2018){Kingma} and {Dhariwal}]{Kingma2018}
{Kingma}, D.~P. and {Dhariwal}, P.
\newblock {Glow: Generative Flow with Invertible 1x1 Convolutions}.
\newblock \emph{arXiv e-prints}, art. arXiv:1807.03039, July 2018.

\bibitem[Kriege et~al.(2020)Kriege, Johansson, and Morris]{Kriege2020}
Kriege, N.~M., Johansson, F.~D., and Morris, C.
\newblock A survey on graph kernels.
\newblock \emph{Applied Network Science}, 5\penalty0 (1):\penalty0 1--42, 2020.

\bibitem[{Li} et~al.(2019){Li}, {Vogelsberger}, {Marinacci}, and
  {Gnedin}]{Li2019}
{Li}, H., {Vogelsberger}, M., {Marinacci}, F., and {Gnedin}, O.~Y.
\newblock {Disruption of giant molecular clouds and formation of bound star
  clusters under the influence of momentum stellar feedback}.
\newblock \emph{\mnras}, 487\penalty0 (1):\penalty0 364--380, July 2019.
\newblock \doi{10.1093/mnras/stz1271}.

\bibitem[{Luo} et~al.(2015){Luo}, {Zhao}, {Zhao}, {Deng}, {Liu}, {Jing},
  {Wang}, {Zhang}, {Shi}, {Cui}, {Chu}, {Li}, {Bai}, {Wu}, {Cai}, {Cao}, {Cao},
  {Carlin}, {Chen}, {Chen}, {Chen}, {Chen}, {Chen}, {Chen}, {Chen},
  {Christlieb}, {Chu}, {Cui}, {Dong}, {Du}, {Fan}, {Feng}, {Fu}, {Gao}, {Gong},
  {Gu}, {Guo}, {Han}, {He}, {Hou}, {Hou}, {Hou}, {Hu}, {Hu}, {Hu}, {Huo},
  {Jia}, {Jiang}, {Jiang}, {Jiang}, {Jin}, {Kong}, {Kong}, {Lei}, {Li}, {Li},
  {Li}, {Li}, {Li}, {Li}, {Li}, {Li}, {Li}, {Li}, {Li}, {Li}, {Liang}, {Lin},
  {Liu}, {Liu}, {Liu}, {Liu}, {Lu}, {Luo}, {Mao}, {Newberg}, {Ni}, {Qi}, {Qi},
  {Shen}, {Shi}, {Song}, {Song}, {Su}, {Su}, {Tang}, {Tao}, {Tian}, {Wang},
  {Wang}, {Wang}, {Wang}, {Wang}, {Wang}, {Wang}, {Wang}, {Wang}, {Wang},
  {Wang}, {Wang}, {Wang}, {Wang}, {Wang}, {Wang}, {Wang}, {Wang}, {Wang},
  {Wang}, {Wei}, {Wei}, {Wu}, {Wu}, {Wu}, {Wu}, {Xing}, {Xu}, {Xu}, {Xu},
  {Yan}, {Yang}, {Yang}, {Yang}, {Yang}, {Yao}, {Yu}, {Yuan}, {Yuan}, {Yuan},
  {Yuan}, {Zhai}, {Zhang}, {Zhang}, {Zhang}, {Zhang}, {Zhang}, {Zhang},
  {Zhang}, {Zhang}, {Zhao}, {Zhou}, {Zhou}, {Zhu}, {Zhu}, {Zou}, and
  {Zuo}]{Luo2015}
{Luo}, A.~L., {Zhao}, Y.-H., {Zhao}, G., {Deng}, L.-C., {Liu}, X.-W., {Jing},
  Y.-P., {Wang}, G., {Zhang}, H.-T., {Shi}, J.-R., {Cui}, X.-Q., {Chu}, Y.-Q.,
  {Li}, G.-P., {Bai}, Z.-R., {Wu}, Y., {Cai}, Y., {Cao}, S.-Y., {Cao}, Z.-H.,
  {Carlin}, J.~L., {Chen}, H.-Y., {Chen}, J.-J., {Chen}, K.-X., {Chen}, L.,
  {Chen}, X.-L., {Chen}, X.-Y., {Chen}, Y., {Christlieb}, N., {Chu}, J.-R.,
  {Cui}, C.-Z., {Dong}, Y.-Q., {Du}, B., {Fan}, D.-W., {Feng}, L., {Fu}, J.-N.,
  {Gao}, P., {Gong}, X.-F., {Gu}, B.-Z., {Guo}, Y.-X., {Han}, Z.-W., {He},
  B.-L., {Hou}, J.-L., {Hou}, Y.-H., {Hou}, W., {Hu}, H.-Z., {Hu}, N.-S., {Hu},
  Z.-W., {Huo}, Z.-Y., {Jia}, L., {Jiang}, F.-H., {Jiang}, X., {Jiang}, Z.-B.,
  {Jin}, G., {Kong}, X., {Kong}, X., {Lei}, Y.-J., {Li}, A.-H., {Li}, C.-H.,
  {Li}, G.-W., {Li}, H.-N., {Li}, J., {Li}, Q., {Li}, S., {Li}, S.-S., {Li},
  X.-N., {Li}, Y., {Li}, Y.-B., {Li}, Y.-P., {Liang}, Y., {Lin}, C.-C., {Liu},
  C., {Liu}, G.-R., {Liu}, G.-Q., {Liu}, Z.-G., {Lu}, W.-Z., {Luo}, Y., {Mao},
  Y.-D., {Newberg}, H., {Ni}, J.-J., {Qi}, Z.-X., {Qi}, Y.-J., {Shen}, S.-Y.,
  {Shi}, H.-M., {Song}, J., {Song}, Y.-H., {Su}, D.-Q., {Su}, H.-J., {Tang},
  Z.-H., {Tao}, Q.-S., {Tian}, Y., {Wang}, D., {Wang}, D.-Q., {Wang}, F.-F.,
  {Wang}, G.-M., {Wang}, H., {Wang}, H.-C., {Wang}, J., {Wang}, J.-N., {Wang},
  J.-L., {Wang}, J.-P., {Wang}, J.-X., {Wang}, L., {Wang}, M.-X., {Wang},
  S.-G., {Wang}, S.-Q., {Wang}, X., {Wang}, Y.-N., {Wang}, Y., {Wang}, Y.-F.,
  {Wang}, Y.-F., {Wei}, P., {Wei}, M.-Z., {Wu}, H., {Wu}, K.-F., {Wu}, X.-B.,
  {Wu}, Y.-Z., {Xing}, X.-Z., {Xu}, L.-Z., {Xu}, X.-Q., {Xu}, Y., {Yan}, T.-S.,
  {Yang}, D.-H., {Yang}, H.-F., {Yang}, H.-Q., {Yang}, M., {Yao}, Z.-Q., {Yu},
  Y., {Yuan}, H., {Yuan}, H.-B., {Yuan}, H.-L., {Yuan}, W.-M., {Zhai}, C.,
  {Zhang}, E.-P., {Zhang}, H.-W., {Zhang}, J.-N., {Zhang}, L.-P., {Zhang}, W.,
  {Zhang}, Y., {Zhang}, Y.-X., {Zhang}, Z.-C., {Zhao}, M., {Zhou}, F., {Zhou},
  X., {Zhu}, J., {Zhu}, Y.-T., {Zou}, S.-C., and {Zuo}, F.
\newblock {The first data release (DR1) of the LAMOST regular survey}.
\newblock \emph{Research in Astronomy and Astrophysics}, 15\penalty0
  (8):\penalty0 1095, August 2015.
\newblock \doi{10.1088/1674-4527/15/8/002}.

\bibitem[{Majewski} et~al.(2017){Majewski}, {Schiavon}, {Frinchaboy}, {Allende
  Prieto}, {Barkhouser}, {Bizyaev}, {Blank}, {Brunner}, {Burton}, {Carrera},
  {Chojnowski}, {Cunha}, {Epstein}, {Fitzgerald}, {Garc{\'{\i}}a P{\'e}rez},
  {Hearty}, {Henderson}, {Holtzman}, {Johnson}, {Lam}, {Lawler}, {Maseman},
  {M{\'e}sz{\'a}ros}, {Nelson}, {Nguyen}, {Nidever}, {Pinsonneault},
  {Shetrone}, {Smee}, {Smith}, {Stolberg}, {Skrutskie}, {Walker}, {Wilson},
  {Zasowski}, {Anders}, {Basu}, {Beland}, {Blanton}, {Bovy}, {Brownstein},
  {Carlberg}, {Chaplin}, {Chiappini}, {Eisenstein}, {Elsworth}, {Feuillet},
  {Fleming}, {Galbraith-Frew}, {Garc{\'{\i}}a}, {Garc{\'{\i}}a-Hern{\'a}ndez},
  {Gillespie}, {Girardi}, {Gunn}, {Hasselquist}, {Hayden}, {Hekker}, {Ivans},
  {Kinemuchi}, {Klaene}, {Mahadevan}, {Mathur}, {Mosser}, {Muna}, {Munn},
  {Nichol}, {O'Connell}, {Parejko}, {Robin}, {Rocha-Pinto}, {Schultheis},
  {Serenelli}, {Shane}, {Silva Aguirre}, {Sobeck}, {Thompson}, {Troup},
  {Weinberg}, and {Zamora}]{Majewski2017}
{Majewski}, S.~R., {Schiavon}, R.~P., {Frinchaboy}, P.~M., {Allende Prieto},
  C., {Barkhouser}, R., {Bizyaev}, D., {Blank}, B., {Brunner}, S., {Burton},
  A., {Carrera}, R., {Chojnowski}, S.~D., {Cunha}, K., {Epstein}, C.,
  {Fitzgerald}, G., {Garc{\'{\i}}a P{\'e}rez}, A.~E., {Hearty}, F.~R.,
  {Henderson}, C., {Holtzman}, J.~A., {Johnson}, J.~A., {Lam}, C.~R., {Lawler},
  J.~E., {Maseman}, P., {M{\'e}sz{\'a}ros}, S., {Nelson}, M., {Nguyen}, D.~C.,
  {Nidever}, D.~L., {Pinsonneault}, M., {Shetrone}, M., {Smee}, S., {Smith},
  V.~V., {Stolberg}, T., {Skrutskie}, M.~F., {Walker}, E., {Wilson}, J.~C.,
  {Zasowski}, G., {Anders}, F., {Basu}, S., {Beland}, S., {Blanton}, M.~R.,
  {Bovy}, J., {Brownstein}, J.~R., {Carlberg}, J., {Chaplin}, W., {Chiappini},
  C., {Eisenstein}, D.~J., {Elsworth}, Y., {Feuillet}, D., {Fleming}, S.~W.,
  {Galbraith-Frew}, J., {Garc{\'{\i}}a}, R.~A., {Garc{\'{\i}}a-Hern{\'a}ndez},
  D.~A., {Gillespie}, B.~A., {Girardi}, L., {Gunn}, J.~E., {Hasselquist}, S.,
  {Hayden}, M.~R., {Hekker}, S., {Ivans}, I., {Kinemuchi}, K., {Klaene}, M.,
  {Mahadevan}, S., {Mathur}, S., {Mosser}, B., {Muna}, D., {Munn}, J.~A.,
  {Nichol}, R.~C., {O'Connell}, R.~W., {Parejko}, J.~K., {Robin}, A.~C.,
  {Rocha-Pinto}, H., {Schultheis}, M., {Serenelli}, A.~M., {Shane}, N., {Silva
  Aguirre}, V., {Sobeck}, J.~S., {Thompson}, B., {Troup}, N.~W., {Weinberg},
  D.~H., and {Zamora}, O.
\newblock {The Apache Point Observatory Galactic Evolution Experiment
  (APOGEE)}.
\newblock \emph{\aj}, 154:\penalty0 94, September 2017.
\newblock \doi{10.3847/1538-3881/aa784d}.

\bibitem[{Mok} et~al.(2019){Mok}, {Chandar}, and {Fall}]{Mok2019}
{Mok}, A., {Chandar}, R., and {Fall}, S.~M.
\newblock {Constraints on Upper Cutoffs in the Mass Functions of Young Star
  Clusters}.
\newblock \emph{\apj}, 872\penalty0 (1):\penalty0 93, February 2019.
\newblock \doi{10.3847/1538-4357/aaf6ea}.

\bibitem[Morris et~al.(2017)Morris, Kersting, and Mutzel]{Morris2017}
Morris, C., Kersting, K., and Mutzel, P.
\newblock Glocalized weisfeiler-lehman graph kernels: Global-local feature maps
  of graphs.
\newblock In \emph{2017 IEEE International Conference on Data Mining (ICDM)},
  pp.\  327--336. IEEE, 2017.

\bibitem[{Ness} et~al.(2018){Ness}, {Rix}, {Hogg}, {Casey}, {Holtzman},
  {Fouesneau}, {Zasowski}, {Geisler}, {Shetrone}, {Minniti}, {Frinchaboy}, and
  {Roman-Lopes}]{Ness2018}
{Ness}, M., {Rix}, H.~W., {Hogg}, D.~W., {Casey}, A.~R., {Holtzman}, J.,
  {Fouesneau}, M., {Zasowski}, G., {Geisler}, D., {Shetrone}, M., {Minniti},
  D., {Frinchaboy}, P.~M., and {Roman-Lopes}, A.
\newblock {Galactic Doppelg{\"a}ngers: The Chemical Similarity Among Field
  Stars and Among Stars with a Common Birth Origin}.
\newblock \emph{\apj}, 853\penalty0 (2):\penalty0 198, February 2018.
\newblock \doi{10.3847/1538-4357/aa9d8e}.

\bibitem[{Price-Jones} \& {Bovy}(2018){Price-Jones} and
  {Bovy}]{Price-Jones2018}
{Price-Jones}, N. and {Bovy}, J.
\newblock {The dimensionality of stellar chemical space using spectra from the
  Apache Point Observatory Galactic Evolution Experiment}.
\newblock \emph{\mnras}, 475\penalty0 (1):\penalty0 1410--1425, March 2018.
\newblock \doi{10.1093/mnras/stx3198}.

\bibitem[Schulz et~al.(2022)Schulz, Horv{\'a}th, Welke, and Wrobel]{Schulz2022}
Schulz, T.~H., Horv{\'a}th, T., Welke, P., and Wrobel, S.
\newblock A generalized weisfeiler-lehman graph kernel.
\newblock \emph{Machine Learning}, 111\penalty0 (7):\penalty0 2601--2629, 2022.

\bibitem[Shervashidze et~al.(2011)Shervashidze, Schweitzer, van Leeuwen,
  Mehlhorn, and Borgwardt]{Shervashidze2011}
Shervashidze, N., Schweitzer, P., van Leeuwen, E.~J., Mehlhorn, K., and
  Borgwardt, K.~M.
\newblock Weisfeiler-lehman graph kernels.
\newblock \emph{J. Mach. Learn. Res.}, 12:\penalty0 2539--2561, 2011.

\bibitem[{Smith} et~al.(2022){Smith}, {Dale}, {Jaffa}, and {Krause}]{Smith2022}
{Smith}, J.~D., {Dale}, J.~E., {Jaffa}, S.~E., and {Krause}, M. G.~H.
\newblock {Star cluster formation in clouds with externally driven turbulence}.
\newblock \emph{\mnras}, 516\penalty0 (3):\penalty0 4212--4219, November 2022.
\newblock \doi{10.1093/mnras/stac2295}.

\bibitem[{Spina} et~al.(2022){Spina}, {Magrini}, {Sacco}, {Casali},
  {Vallenari}, {Tautvai{\v{s}}ien{\.{e}}}, {Jim{\'e}nez-Esteban}, {Gilmore},
  {Randich}, {Feltzing}, {Jeffries}, {Bensby}, {Bragaglia}, {Smiljanic},
  {Carraro}, {Morbidelli}, and {Zaggia}]{Spina2022}
{Spina}, L., {Magrini}, L., {Sacco}, G.~G., {Casali}, G., {Vallenari}, A.,
  {Tautvai{\v{s}}ien{\.{e}}}, G., {Jim{\'e}nez-Esteban}, F., {Gilmore}, G.,
  {Randich}, S., {Feltzing}, S., {Jeffries}, R.~D., {Bensby}, T., {Bragaglia},
  A., {Smiljanic}, R., {Carraro}, G., {Morbidelli}, L., and {Zaggia}, S.
\newblock {The Gaia-ESO Survey: Chemical tagging in the thin disk. Open
  clusters blindly recovered in the elemental abundance space}.
\newblock \emph{\aap}, 668:\penalty0 A16, December 2022.
\newblock \doi{10.1051/0004-6361/202243316}.

\bibitem[{Ting} et~al.(2015){Ting}, {Conroy}, and {Goodman}]{Ting2015}
{Ting}, Y.-S., {Conroy}, C., and {Goodman}, A.
\newblock {Prospects for Chemically Tagging Stars in the Galaxy}.
\newblock \emph{\apj}, 807\penalty0 (1):\penalty0 104, July 2015.
\newblock \doi{10.1088/0004-637X/807/1/104}.

\bibitem[{Ting} et~al.(2016){Ting}, {Conroy}, and {Rix}]{Ting2016}
{Ting}, Y.-S., {Conroy}, C., and {Rix}, H.-W.
\newblock {APOGEE Chemical Tagging Constraint on the Maximum Star Cluster Mass
  in the Alpha-enhanced Galactic Disk}.
\newblock \emph{\apj}, 816\penalty0 (1):\penalty0 10, January 2016.
\newblock \doi{10.3847/0004-637X/816/1/10}.

\bibitem[{Vincenzo} et~al.(2021){Vincenzo}, {Weinberg}, {Miglio}, {Lane}, and
  {Roman-Lopes}]{Vincenzo2021}
{Vincenzo}, F., {Weinberg}, D.~H., {Miglio}, A., {Lane}, R.~R., and
  {Roman-Lopes}, A.
\newblock {The distribution of [$\alpha$/Fe] in the Milky Way disc}.
\newblock \emph{arXiv e-prints}, art. arXiv:2101.04488, January 2021.

\bibitem[{Weinberg} et~al.(2019){Weinberg}, {Holtzman}, {Hasselquist}, {Bird},
  {Johnson}, {Shetrone}, {Sobeck}, {Allende Prieto}, {Bizyaev}, {Carrera},
  {Cohen}, {Cunha}, {Ebelke}, {Fernandez-Trincado},
  {Garc{\'\i}a-Hern{\'a}ndez}, {Hayes}, {J{\"o}nsson}, {Lane}, {Majewski},
  {Malanushenko}, {M{\'e}sz{\'a}ros}, {Nidever}, {Nitschelm}, {Pan}, {Rix},
  {Rybizki}, {Schiavon}, {Schneider}, {Wilson}, and {Zamora}]{Weinberg2019}
{Weinberg}, D.~H., {Holtzman}, J.~A., {Hasselquist}, S., {Bird}, J.~C.,
  {Johnson}, J.~A., {Shetrone}, M., {Sobeck}, J., {Allende Prieto}, C.,
  {Bizyaev}, D., {Carrera}, R., {Cohen}, R.~E., {Cunha}, K., {Ebelke}, G.,
  {Fernandez-Trincado}, J.~G., {Garc{\'\i}a-Hern{\'a}ndez}, D.~A., {Hayes},
  C.~R., {J{\"o}nsson}, H., {Lane}, R.~R., {Majewski}, S.~R., {Malanushenko},
  V., {M{\'e}sz{\'a}ros}, S., {Nidever}, D.~L., {Nitschelm}, C., {Pan}, K.,
  {Rix}, H.-W., {Rybizki}, J., {Schiavon}, R.~P., {Schneider}, D.~P., {Wilson},
  J.~C., and {Zamora}, O.
\newblock {Chemical Cartography with APOGEE: Multi-element Abundance Ratios}.
\newblock \emph{\apj}, 874\penalty0 (1):\penalty0 102, March 2019.
\newblock \doi{10.3847/1538-4357/ab07c7}.

\end{thebibliography}
\bibliographystyle{icml2023}
\end{document}